\documentclass[3p,twocolumn]{elsarticle}

\usepackage{graphicx}
\usepackage{epsfig}
\usepackage{amsmath}

\begin{document}

\title{Ripples and dots generated by lattice gases}

\author{G\'eza \'Odor (1), Bartosz Liedke, Karl-Heinz Heinig,
and Jeffrey Kelling (2)}

\address{(1) Research Institute for Technical Physics and
Materials Science, 
P.O.Box 49, H-1525 Budapest, Hungary\\
(2) Institute of Ion Beam Physics and Materials Research
Helmholtz-Zentrum Dresden-Rossendorf 
P.O.Box 51 01 19, 01314 Dresden, Germany}    

\begin{abstract}
We show that the emergence of different surface patterns (ripples, dots) 
can be well understood by a suitable mapping onto the simplest 
nonequilibrium lattice gases and cellular automata.
Using this efficient approach difficult, unanswered questions
of surface growth and its scaling can be studied. 
The mapping onto binary variables facilitates effective simulations 
and enables one to consider very large system sizes. 
We have confirmed that the fundamental Kardar-Parisi-Zhang (KPZ) 
universality class is stable against a competing roughening diffusion, 
while a strong smoothing diffusion leads to logarithmic growth, a
mean-field type behavior in two dimensions. 
The model can also describe anisotropic surface diffusion processes
effectively. By analyzing the time-dependent structure factor we
give numerical estimates for the wavelength coarsening behavior.
\end{abstract}
\maketitle

\section{Introduction}

In nanotechnologies, patterns of large areas are to be assembled in
a cost effective way, possibly by self-organization processes.
Different materials under different conditions 
have been shown to exhibit dot and ripple formation in a rather universal 
manner \cite{FacskoS}.
Universality appears, when microscopic details of interactions
become irrelevant, i.e. the diverging correlation length and the complex
behavior of the system dominates. This is typical in nonequilibrium systems
with currents \cite{S-Z,Orev}.
It was pointed out \cite{RLP90,krug99} that grooved surfaces and growth
instabilities may emerge as the consequence of broken detailed balance
condition 
\begin{equation}\label{dbal}
P(C) R_{C\to C'} \ne P(C') R_{C'\to C}
\end{equation}
where $P(C)$ denotes the probability of the state $C$ and $R_{C\to C'}$ 
is the transition rate between states $C$ and $C'$.
This means that complex structures and patterns can emerge during relaxation
in nonequilibrium system. In case of ion-beam-induced modification of surfaces
the first theory to explain pattern formation was suggested long time ago
\cite{BH88}.

To demonstrate the general tendency of self-organization
we present here simple case studies of fundamental models of statistical 
physics, showing universal scaling and pattern formation.
Understanding of basic surface growth models still lacks many details,
although the basic universality classes and important models
have been introduced \cite{barabasi,MCB02}.
Analytical tools like continuum field methods have limited 
applicability, while numerical simulations have achieved limited precision.

One of the simplest nonlinear stochastic differential equation
set up by Kardar, Parisi and Zhang (KPZ) \cite{KPZeq} describes 
the dynamics of surface growth processes in the thermodynamic limit.
It specifies the evolution of the height function $h(\mathbf{x},t)$ in the
$d$-dimensional space
\begin{equation}  \label{KPZ-e}
\partial_t h(\mathbf{x},t) = v + \sigma\nabla^2 h(\mathbf{x},t) +
\lambda(\nabla h(\mathbf{x},t))^2 + \eta(\mathbf{x},t) \ .
\end{equation}
Here $v$ and $\lambda$ are the amplitudes of the mean and local growth
velocity, $\sigma$ can be understood as a coefficient of
surface tension driven smoothing and $\eta$ roughens the surface by 
a zero-average, Gaussian noise field exhibiting the variance:
$\langle\eta(\mathbf{x},t)\eta(\mathbf{x^{\prime}},t^{\prime})\rangle = 2 D
\delta^d (\mathbf{x-x^{\prime}})(t-t^{\prime})$.
The notation $D$ is used for the noise amplitude and $\langle\rangle$
means the distribution average. The surface scaling exponents are known 
in $\left( 1+1\right) d$ \cite{kardar87}, but for the case
considered here, 2-dimensional surfaces of 3-dimensional system, as well
as in higher dimensions approximations are available only. 
The existence of a finite upper critical dimension, where a 
smooth mean-field behavior would enter is also debated.
As the result of the competition of roughening and smoothing terms,
models described by the KPZ equation exhibit a roughening phase transition
between a weak-coupling regime ($\lambda <\lambda _{c}$), governed by the
Edwards-Wilkinson (EW) fixed point at $\lambda =0$ \cite{EWc}, and a strong
coupling phase. The strong coupling fixed point is inaccessible by the
perturbative renormalization group method.
Therefore, the KPZ phase space has been the subject of controversies
for a long time.

Mapping of surface growth onto reaction-diffusion system allows effective 
numerical simulations and understanding of basic universality classes 
\cite{dimerlcikk,Obook08}.
In one dimension a discrete, restricted solid on solid (RSOS) realization 
of the KPZ growth is equivalent to the Asymmetric Simple Exclusion 
Process of particles \cite{Rost81}, while some of us
have shown that the roof-top model mapping \cite{kpz-asepmap,meakin} 
can be generalized to higher dimensions \cite{asep2dcikk,asepddcikk}.

\begin{figure}[ht]
\begin{center}
\epsfxsize=70mm
\epsffile{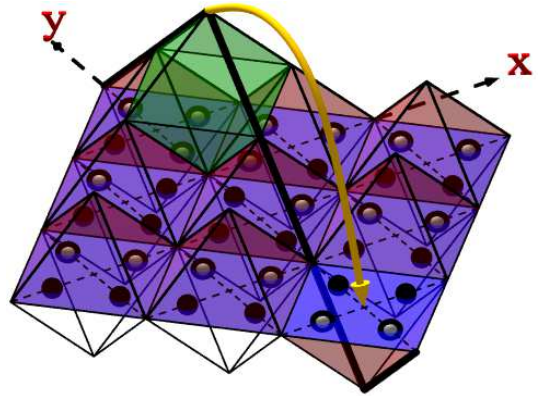}
\caption{Mapping of the $2+1$ dimensional surface
growth onto the $2d$ particle model (bullets).
Surface attachment (with probability $p$) and
detachment  (with probability $q$) corresponds to
exchanges of particles, or to anisotropic
diffusion of dimers in the bisectrix direction
of the $x$ and $y$ axes. The crossing points of dashed lines
show the base sub-lattice to be updated. Thick solid line on
the surface shows the $y$ cross-section as a reminder for the
one-dimensional roof-top model.
When the shown desorption/absorption steps are executed simultaneously
they realize a surface diffusion jump of distance $s=3$ 
along the $y$ axis.}
\label{2dM}
\end{center}
\end{figure}

In two dimensions one can map KPZ processes onto anisotropic, 
but oriented migration of directed dimers \cite{asep2dcikk}. 
This mapping is interesting not conceptually only, linking 
nonequilibrium surface growth with the dynamics of driven 
lattice gases \cite{S-Z,DickMar}, but provides an efficient 
tool for investigating debated and unresolved problems numerically
\cite{asepddcikk}. 
The surface built up from the octahedra can be represented by the
edges meeting in the up/down middle vertexes (see Fig.~\ref{2dM}).
The up edges in the $\chi=x$ or $\chi=y$ directions at the lattice site
$i,j$ are represented by $\sigma_{\chi}(i,j)=+1$, while the down ones 
by $\sigma_{\chi}(i,j)=-1$ slopes of the model. 
In this way an octahedron site deposition flips four edges, 
meaning two '$+1$'$\leftrightarrow$'$-1$' exchanges:
one in the $x$ and one in the $y$ direction. This can be 
described by the generalized update rule:
\begin{equation}
\left(
\begin{array}{cc}
   -1 & 1 \\
   -1 & 1
\end{array}
\right)\overset{p}{\underset{q}{\rightleftharpoons }}\left(
\begin{array}{cc}
   1 & -1 \\
   1 & -1
\end{array}
\right) \label{rule}
\end{equation}
with probability $p$ for attachment and probability $q$ for
detachment.
We can also call '$+1$'-s as particles and '$-1$'-s as holes of
a reaction-diffusion model on a square lattice. 
Thus, an attachment/detachment update corresponds to a single step 
motion of an oriented dimer in the bisectrix direction of the 
$x$ and $y$ axes. We update the neighborhood of the sub-lattice
points, which are the crossing-points of the dashed lines.
In \cite{asep2dcikk,asepddcikk} we derived how this mapping 
connects the microscopic model to the KPZ equation 
with $\lambda = 2\frac{p}{p+q}-1$ and investigated its 
surface scaling numerically. In particular we have shown that
for the $\lambda=0$ case our simulation results agree with 
those of the exactly known EW class.

Surface diffusion is a much studied basic process \cite{krug-rev}.
Several atomistic models have been constructed and investigated with the
aim of realizing Mullins-Herring (MH) diffusion \cite{herring50,mullins}
and scaling (for a recent review see \cite{MBEpers}).
The Langevin equation of MH diffusion is a linear one, with a 
$\nabla^4$ lowest order gradient term
\begin{equation}
\label{MH0e}
\partial_t h({\mathbf x},t)  = -\kappa\nabla^4 h({\mathbf x},t)  + \eta({\mathbf x},t) \ .
\end{equation}
resulting from a curvature driven surface current.
This equation is exactly solvable and exhibits scale invariance
of the roughness. 
In Sect.~\ref{gensect} we show how this can be realized 
using the octahedron model.

\subsection{The simulations}

Dynamic, bit-coded simulations were run on conserved lattice gas models
of sizes $L\times L$ following the rule (\ref{rule}). 
The surface heights were reconstructed from the slopes
\begin{equation}
h_{i,j} = \sum_{l=1}^i \sigma_x(l,1) + \sum_{k=1}^j \sigma_y(i,k) \ .
\end{equation}
The interface width defined as
\begin{equation}
w(L,t) =
\Bigl[
\frac{1}{L^2} \, \sum_{i,j}^L \,h^2_{i,j}(t)  -
\Bigl(\frac{1}{L^2} \, \sum_{i,j}^L \,h_{i,j}(t) \Bigr)^2
\Bigr]^{1/2} 
\end{equation}
has been calculated at certain sampling times $t$.
In the absence of any characteristic length, growth processes 
are expected to follow a power-law behavior and the average surface 
width $W=\overline w$ can be described by the 
{\em Family-Vicsek}~\cite{family} scaling law:
\begin{equation}
\label{FV-forf}
W(L,t) \simeq L^{\alpha} f(t / L^z),
\end{equation}
with the universal scaling function $f(u)$ of the form:
\begin{equation}
\label{FV-fu}
f(u) \sim
\left\{ \begin{array}{lcl}
     u^{\beta}     & {\rm if} & u \ll 1 \\
     {\rm const.} & {\rm if} & u \gg 1 \ .
\end{array}
\right .
\end{equation}
Here $\alpha$ is the roughness exponent of the stationary regime,
when the correlation length has exceeded the system size $L$ and
$\beta$ is the growth exponent, describing the intermediate time 
behavior. The dynamical exponent $z$ is the ratio
\begin{equation}\label{zlaw}
z = \alpha/\beta \ .
\end{equation}

For the 2d KPZ universality class our numerical results
\cite{asepddcikk}
\begin{equation}\label{KPZclass}
\alpha=0.39(1), \ \ \ \beta=0.245(5), \ \ \ z=1.6(1) 
\end{equation}
are in agreement with most of the literature values \cite{MPP,AA04}.
The MH class, with non-conservative noise is characterized 
by the exponents
\begin{equation}\label{MHclass}
\alpha=1, \ \ \ \beta=1/4, \ \ \ z=4 
\end{equation}
in two dimensions, while in case of conservative 
(diffusive) noise the the exponents are
\begin{equation}\label{MHCclass}
\alpha=0, \ \ \ \beta=0, \ \ \ z=4 
\end{equation}
and the growth can be logarithmic \cite{barabasi}.
Note that in both cases the the dynamics is very slow,
described by the large dynamical exponent $z=4$, which makes 
the simulations time consuming.

Emerging patterns can be characterized by the time dependent
power spectrum density (PSD) of the interface 
\begin{equation}\label{PSD}
S({\mathbf k},t) = \langle h({\mathbf k},t) h(-{\mathbf k},t) \rangle  \ ,
\end{equation} 
where the height in the Fourier space is computed as
\begin{equation} \label{hF}
h({\mathbf k},t) = \frac{1}{L^{d/2}} \sum_{{\mathbf\chi}} [ h({\mathbf\chi},t) - \overline h ] \ {\rm exp}(i {\mathbf k \chi}) \ .
\end{equation}
We computed $h({\mathbf k},t)$ from the surface profiles with the FFT method
and determined $S({\mathbf k},t)$ by averaging over $x$ and $y$ directions 
in case of isotropic patterns and only over the direction
perpendicular to the ripples in case of $x/y$ anisotropy.
We have also calculated the characteristic wavelength 
of the patterns from the maxima of $S({\mathbf k},t)$.

The simulations were started from a flat surface, corresponding to a 
zig-zag configuration (alternating '1' and '0' lines of the lattice gas)
of the slopes (see Fig.\ref{2dM}) with periodic boundary conditions.
In the model each lattice site can be characterized by $16$ different
local slope configurations, but we update them only when condition
(\ref{rule}) is satisfied. Furthermore, due to the surface continuity,
two slopes become redundant, when considering the neighbors.
Thus we can describe a lattice site by using only two bits.
This permits efficient storage management and large system sizes $L$. 
The updates can be performed by bit operations either on multiple 
samples at once or on multiple sites in parallel. 
Our multi-sample algorithm proved to be $\sim 40$ times faster than the 
conventional FORTRAN 90 code, while a speedup factor of 
$\sim 240$ was measured using the multi-site CUDA code on a 
NVIDIA C2070 graphics card in comparison with a single I5 CPU 
core running on $2.8$ GHz. 
The latter results could be achieved on huge system sizes up to
$L=2^{17}$. More details are published elsewhere 
\cite{asep2dcikk,gputexcikk,2dGPUtobepub}.
 
\subsection{Generalizations of the octahedron model} \label{gensect}

An obvious first step of generalization 
is to combine the deposition and the removal processes, 
creating a conserved dynamics. A simultaneous octahedron
detachment and deposition in the neighborhood can realize an elementary 
diffusion step (see Fig.~\ref{2dM}). 
In \cite{patscalcikk} we described RSOS models with 
$\Delta h=\pm 1$ height restriction, which in the limit 
of weak external noise exhibit MH scaling. 
Microscopic models realizing this behavior are mainly unrestricted
solid on solid (SOS) type, which can provide steep slopes and strong
curvatures. However, it turned out that the asymptotic
universality class of the various limited mobility growth models
is a surprisingly subtle issue and some earlier findings proved
to be incorrect due to pathologically slow crossover and extremely
long transient effects.

In our model the target site is chosen in the $\pm x$ or $\pm y$ direction, 
with the probabilities: $D_{+x}$, $D_{-x}$ or $D_{+y}$, $D_{-y}$ 
respectively (see Fig.\ref{2dM}), thus we can follow {\bf anisotropic
surface diffusion} in principle. Throughout of our studies we normalized the
attempt probabilities. The maximal jump distance was fixed to be
$s_m\le 4$ lattice units following computer experiments. 
\begin{figure}[ht]
\begin{center}
\epsfxsize60mm
\epsffile{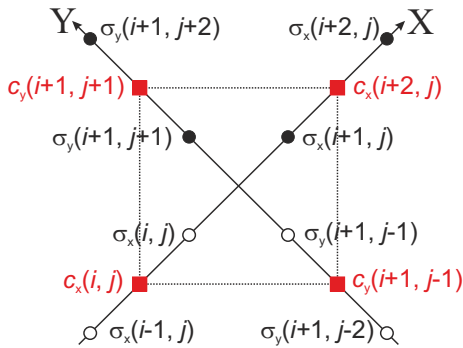}
\caption{Local slope $\sigma_{\chi}(i,j)$ (circles) and curvature
$c_{\chi}(i,j)$ (squares) variables at an update site.
Filled circles correspond to upward, empty ones to downward
slopes of the surface. This plaquette configuration
models a valley bottom site.}
\label{MHupdate}
\end{center}
\end{figure}
To control the surface diffusion we imposed additional constraints
for the accepting a move. We have tried two kinds of rules based on
the local neighborhood configurations. The first one requires that the
height of a particle at the final state is higher than that of its
initial site
\begin{equation}\label{Rproc}
h_\mathrm{fin}-h_\mathrm{ini} \ge 0 \ ,
\end{equation}
which makes the surface more rough (inverse or roughening diffusion).
For details of this so-called Larger Height Octahedron Model (LHOD)
see \cite{patscalcikk}. Note, that LHOD introduces up-down anisotropy,
which leads to a $\nabla^2[\nabla(h)^2]$ type of nonlinearity 
and Molecular Beam Epitaxy (MBE) class scaling \cite{patscalcikk} in general.

\begin{figure}[ht]
\begin{center}
\epsfxsize70mm
\epsffile{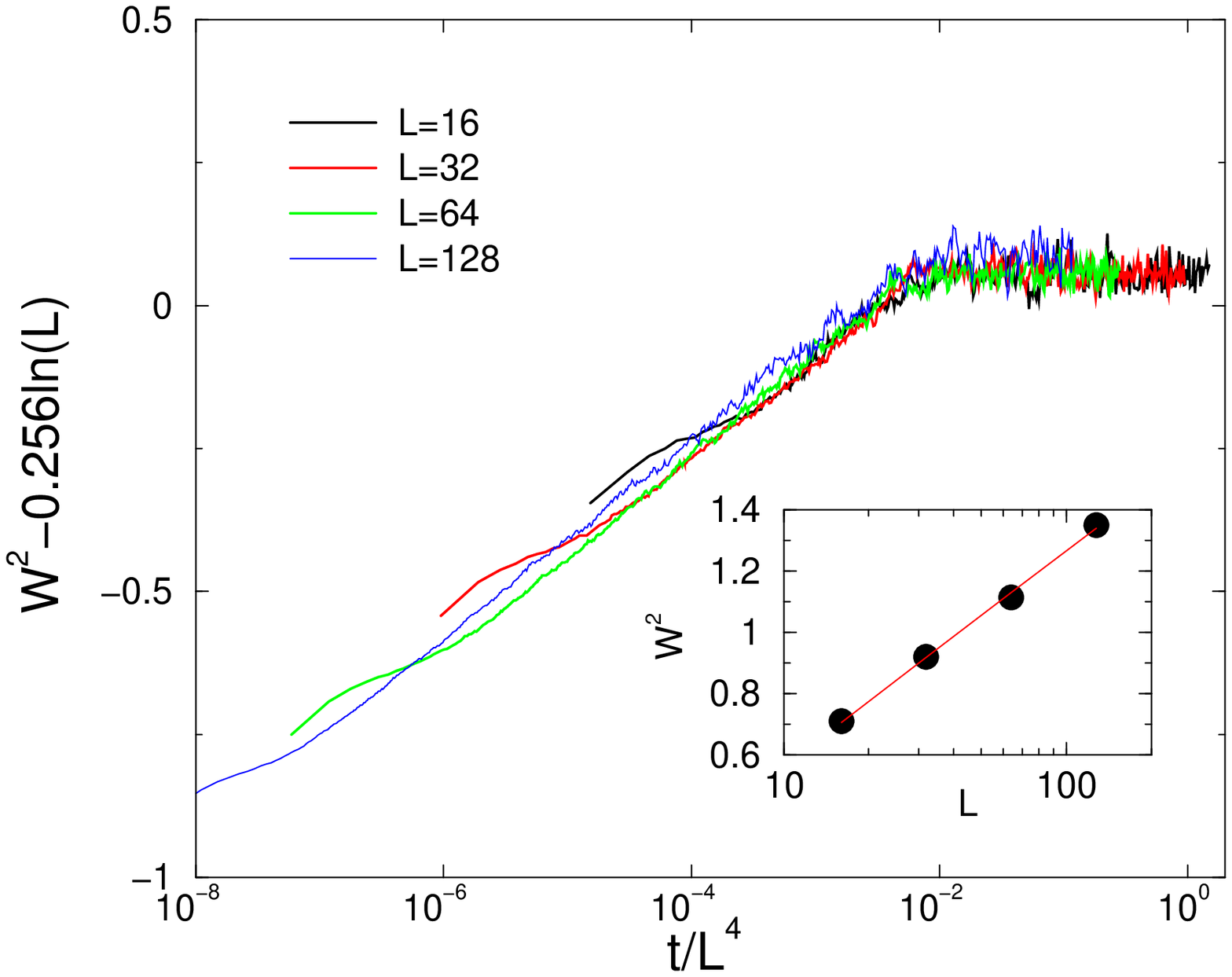}
\caption{Scaling behavior of the isotropic LCOD model for $L=16,32,64,128$
(top to bottom at the left side). The data collapse ($f(u)$) has been
achieved with the MH class exponents with conservative noise(\ref{MHCclass}).
The insert shows the finite size (logarithmic) scaling in the steady state.}
\label{MHN}
\end{center}
\end{figure}

The second one is the so-called Larger Curvature Octahedron Model (LCOD),
which satisfies the detailed balance condition (\ref{dbal}), 
enabling us to realize linear, equilibrium MH diffusion steps.
The local curvature of the surface is calculated at the $4$
edges of squares of the projected octahedra as shown on 
Fig.~\ref{MHupdate}. 
We followed the scaling behavior of our LCOD in case of
smoothing reactions, corresponding to $\kappa > 0$ 
in Eq.~\ref{MH0e}. Due to the purely diffusive noise the growth 
dynamics is logarithmically slow and we find the emergence of the
class (\ref{MHCclass}) scaling (see Figure~\ref{MHN}).

\begin{figure}[ht]
\begin{center}
\epsfxsize90mm
\epsffile{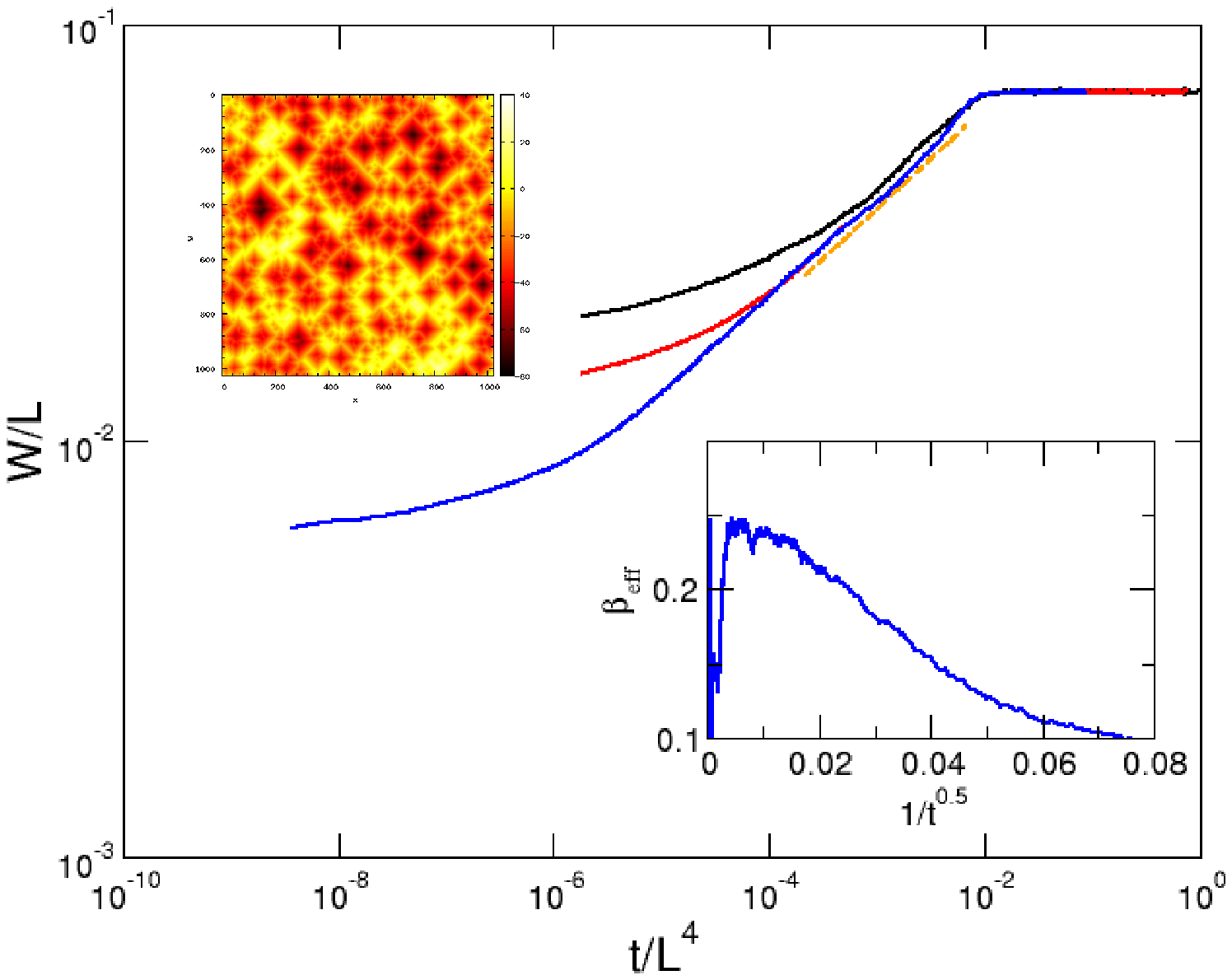}
\caption{Scaling behavior of the isotropic LCOD model for $L=32,64,128$ 
(top to bottom at the left side). The data collapse ($f(u)$) has been 
achieved with the MH class exponents (\ref{MHclass}). 
For the growth exponent fitting (dashed line) results in $\beta=0.25(1)$. 
The right insert shows the same via the local slopes (\ref{beff}).
The left insert is a snapshot of surface heights of the dot patterns 
generated by the LHOD model with parameters: $D_{\pm x}=D_{\pm y}=1$
and $p=q=0.1$ at $t=10^4$ MCs}
\label{MHB}
\end{center}
\end{figure}

If the accepted move increases the local curvature, we call it inverse 
MH (iMH) process. In the continuum model this should correspond to 
$\kappa <0$, an unstable growth, however in the finite, lattice model
formation of pyramid-like structures occurs similarly as in the "$n=2$" 
SOS model \cite{SP94}.
When we simulated the evolution of the iMH process (LCOD model)
in the presence of a small, competing EW type of noise ($p=q=0.05$)
starting from flat initial condition we found a scaling behavior, 
which agrees well with that of MH universality class 
(\ref{MHclass}) (see Fig.~\ref{MHB}).
The effective growth exponent defined as
\begin{equation}  \label{beff}
\beta_\mathrm{eff}(t) = \frac {\ln W(t,L\to\infty) - \ln W(t/2,L\to\infty)} 
{\ln(t) - \ln(t/2)} 
\end{equation}
converges to $\beta=0.25(1)$ before reaching the saturation regime.

\section{Pattern generation by competing inverse MH and KPZ processes}

A further step in generalizing our basic surface models was
the combination of different sub-processes, resulting in 
nonequilibrium system with patterns. 
For example the mixed application of iMH diffusion steps with 
KPZ updates can model a noisy Kuramoto-Sivashinsky (KS) type of equation  
\cite{ks:1977,sivashinsky:1979,KSC}, the inverse KS (iKS):
\begin{equation}
\label{iKS-e}
\partial_t h = \sigma\nabla^2 h + \kappa\nabla^4 h + \lambda(\nabla h)^2 
+ \eta \ .
\end{equation}
However the signs of couplings are reversed with respect to 
the KS equation \footnote{Where both the $\nabla^2$ and the $\nabla^4$ 
terms have negative coefficients and a positive $(\nabla h)^2$
can stabilize the solution.}.
This leads to unstable solutions, the long wave behavior of iKS
is not defined asymptotically in the continuum model. However, in the
lattice regularized version, for intermediate times we expect dynamical 
scaling of the KS based on symmetry and renormalization arguments
\cite{barabasi,AVS74}. Since the roughening/smoothing surface moves
correspond to phase separation/mixing of the lattice gas system it is
reasonable to believe that both cases are described by the same 
universal fixed point behavior.

We performed the surface diffusion steps alternately with the 
deposition and the removal processes. 
Thus terms like $\kappa_{x}\partial^4 h$ and $\kappa_{y}\partial^4 h$ 
with $\kappa_{x}\ne \kappa_{y}$ can be modeled. 
We followed the surface roughness and calculated PSD as well as the
wavelength growth of the pattern formation.

\begin{figure}[ht]
\begin{center}
\epsfxsize70mm
\epsffile{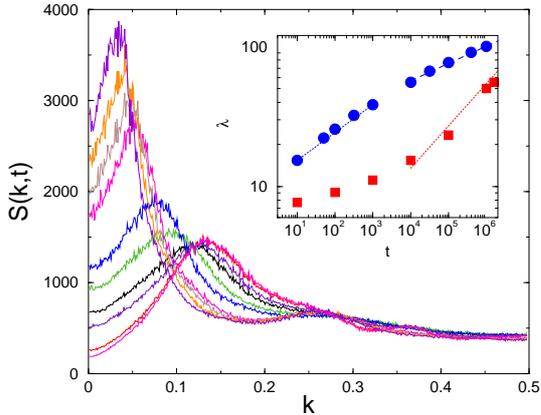}
\caption{PSD for anisotropic diffusion with $D_{y}=1$, $p=q=0.005$ 
for $L=1024$ at different times.
The insert shows the wavelength coarsening calculated from the
maxima of the PSD curves. Squares: LHOD model with fastening 
ripple formation $\lambda\propto t^{0.35(5)}$, 
circles: LCOD model with $p=q=0$. The lines correspond to power-law fits:
$\lambda\propto t^{0.19(1)}$ initially and $\lambda\propto t^{0.12(1)}$
for late times.}
\label{PSDA}
\end{center}
\end{figure}

In case of an anisotropic LHOD a competing EW process always generates 
stable ripple patterns, but if strong up-down anisotropy 
(KPZ instead of EW) is applied for late times the ripples 
become uneven, blurred and cut into smaller pieces.
\begin{figure}[ht]
\begin{center}
\epsfxsize 90mm
\epsffile{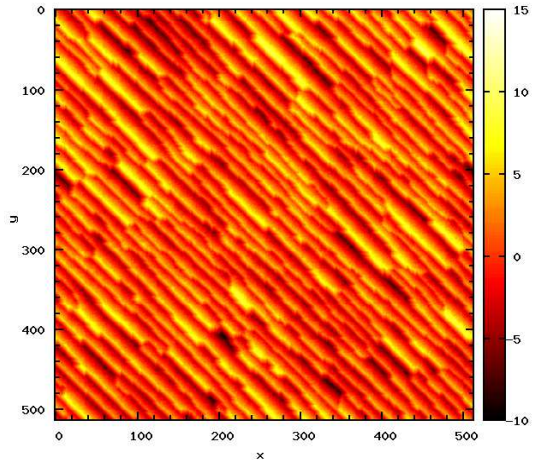}
\caption{Snapshot of surface heights of the ripple patterns generated 
by the parameters $D_x=D_y=0$, $D_{-x}=D_{-y}=1$ (anisotropic, iMH) 
and $p=q=0.05$ adsorption/desorption at $t=10^4$ MCs, in the LHOD model 
of linear size $L=512$. }
\label{AKS}
\end{center}
\end{figure}
If we introduce $x/y$ lattice anisotropy, for example by: $D_x = D_{y} = 0$ 
and $D_{-x} = D_{-y}=1$ the competing weak EW or KPZ moves result 
in {\bf ripple coarsening} (see Fig.~\ref{AKS}), with a wavelength
growth $\lambda\propto t^{0.19(1)}$. We calculated this behavior
from the maxima of the PSD functions. 
However, for late times this coarsening slows down for LCOD 
($\lambda\propto t^{0.12(1)}$) or becomes faster in case of 
nonlinear LHOD ($\lambda\propto t^{0.35(5)}$) as shown on 
Fig.~\ref{PSDA}.
The last value agrees with that of the numerical estimates 
obtained for the two-field model \cite{JCC}.

\begin{figure}[ht]
\begin{center}
\epsfxsize70mm
\epsffile{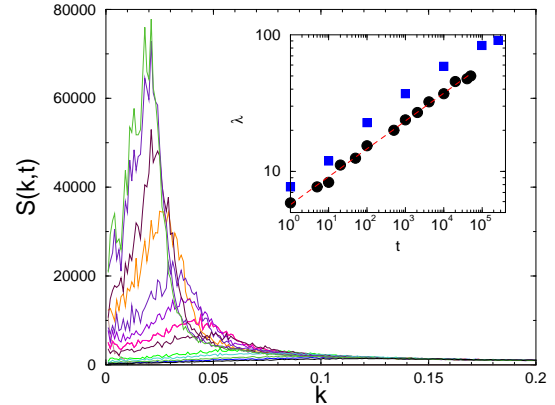}
\caption{PSD in the LCOD model for isotropic surface diffusion 
($D_{\pm x} = D_{\pm y}=1$) with competing EW moves: $p=q=0.005$ 
at different times.
The insert shows the wavelength coarsening calculated from the
maxima of the PSD curves. Circles correspond to the peaks of
the PSD curves, dashed line: power-law fit 
$\lambda\propto t^{0.20(1)}$. 
The boxes show the same for $p=q=0$ in the LCOD. 
For late times the slope of $\lambda(t)$ decreases.}
\label{KSB}
\end{center}
\end{figure}

When isotropic, iMH competes with a weak EW process one can observe 
dot formation in both the LHOD and LCOD model. 
Again, for strong particle deposition or removal the patterns fade away 
for long times and the KPZ scaling emerges.
The insert of Fig.~\ref{MHB} shows a snapshot of the {\bf growing dots} 
in the LHOD model in the presence of weak EW processes.
We have determined the PSDs at different times for $L=1024$ 
and deduced the time dependence of the characteristic wavelength
from the maxima of $S({\bf k},t)$ (see Fig.~\ref{KSB}).

Having confirmed that the LCOD model exhibits MH class scaling, 
we introduced a model for the iKS equation (\ref{iKS-e}) as 
the combination of iMH and KPZ moves.
In \cite{patscalcikk} we presented extensive simulations providing 
numerical evidence for an asymptotic KPZ scaling of this model.
Note, that unlike the KS, the iKS equation exhibits unstable 
wavelength amplification. This can be derived in the Fourier space
by a linear stability analysis. However the dynamical 
scaling behavior of the lattice model corresponding to the
iKS equation may be expected to be the same as that of the KS equation
by symmetry arguments.

\section{Conclusions and outlook}

We have returned to some unresolved questions of basic surface growth
phenomena using mappings and extensive computer simulations.
Contrary to many earlier studies we were able to perform numerical 
analysis in $2+1$ dimensions due to the efficient binary lattice gas
representation. An important advantage of our method as compared to 
the analytical approaches is that we can describe anisotropic 
surface diffusion terms $\kappa_{x}\partial^4 h$, 
$\kappa_{y}\partial^4 h$ by our model.

We have shown that in the zero external noise limit RSOS models
can be constructed with short range interactions exhibiting 
MBE or MH type of surface growth behavior. 
For inverse (roughening) diffusion, which increases the unstable growth
of local curvatures, pyramid-like structures can emerge. The size 
of these structures is limited only by $L$, which is not directly 
comparable with experiments. We created these microscopic models 
in order to study them in competition with the non-conserved KPZ processes.
  
By mapping surface models onto lattice gases, our results imply that 
the scaling behavior of the oriented diffusion of dimers (KPZ) is stable 
against the introduction of an attracting force among them. 
However, a strong repulsion force destroys the fluctuations, 
resulting in mean-field class behavior \cite{patscalcikk}.
We found numerical evidence, by surface scaling and probability 
distribution studies, that the iKS model exhibits KPZ scaling in $2+1$ 
dimensions. 
The construction of a KS model within the framework of the 
lattice gas mapping is under way.
Further studies with different boundary conditions can reveal interesting 
connections of the surface tilt to the particle concentration of the 
lattice gas. We emphasize that results for disorder dependence or 
anomalous diffusion \cite{Clong} can easily be transformed between 
these surface and lattice gas models.
Extension of our mapping can also help to solve more detailed models
of ion-beam induced pattern formation \cite{G10,BHL10}, describing
the observed narrow band of unstable wave-lengths or hexagonal
dot formation \cite{FacskoS}. 

We have also determined the time dependent power spectrum 
and deduced the wavelength growth of the patterns. We have investigated 
this for LHOD and LCOD process, with and without spatial anisotropies. 
In case of uniaxial surface diffusion ripple morphologies have been found, 
while for $x/y$ lattice isotropy dot like pattern formation could be traced. 
The wavelength growth is slow for LCOD models.
Usually we found a $\lambda(t) \propto t^{0.19(1)}$ time dependence,
which slows down further at late times for the linear LCOD model, but 
becomes faster in the nonlinear LHOD model. This agrees with the two-field 
model results \cite{JCC}.
In \cite{patscalcikk} we estimated the characteristic length scale
by a different measure, the average length of the longest slopes.
The growth of that scale was found to be even slower, 
except when steady DC current flowed through the system.  
Since that quantity measures an extremal property of the patterns
one can understand that somewhat different results arise from those
of the PSD analysis.

Finally we point out that these models enable efficient, bit-coded, 
stochastic cellular automaton type of simulation of surfaces which 
can be run extremely fast on advanced graphic cards 
\cite{gputexcikk,2dGPUtobepub}. 

\vskip 1.0cm

\noindent
{\bf Acknowledgments:}\\

Support from the Hungarian research fund OTKA (Grant No. T77629),
the bilateral German-Hungarian exchange program DAAD-M\"OB
(50450744, P-M\"OB/854) is acknowledged. 
The authors thank the access to the HUNGRID and
NVIDIA for supporting the project with high-performance
graphics cards within the framework of Professor Partnership.

\end{document}